\documentclass{iaus}
\usepackage{graphicx}

\title[] 
{A first family of stable configurations\\ to model \\ magnetic fields in stellar radiation zones}

\author[V. Duez et al.]   
{Vincent Duez$^1$, Jonathan Braithwaite$^1$
 \and St{\'e}phane Mathis$^2$}

\affiliation{$^1$Argelander Institut f{\"u}r Astronomie, Universit{\"a}t Bonn, Auf dem H{\"u}gel 71, D-53111 Bonn,
Germany, email: {\tt vduez@astro.uni-bonn.de, jonathan@astro.uni-bonn.de} \\[\affilskip]
$^2$Laboratoire AIM, CEA/DSM-CNRS-Universit{\'e} Paris Diderot, IRFU/SAp Centre de
Saclay, F-91191 Gif-sur-Yvette, France, email: {\tt stephane.mathis@cea.fr}}

\pubyear{2010}
\volume{272}  
\pagerange{??--??}
\jname{Active OB stars: structure, evolution, mass loss and critical limits}
\editors{C. Neiner, G. Wade, G. Meynet \& G. Peters, eds.}
\begin{document}

\maketitle

\begin{abstract} 
We conduct 3D magneto-hydrodynamic (MHD) simulations in order to test the stability of the magnetic equilibrium configuration described by Duez \& Mathis (2010). This analytically-derived configuration describes the lowest energy state for a given helicity in a stellar radiation zone.  The necessity of taking into account the non force-free property of the large-scale, global field is here emphasized. We then show that this configuration is stable. It therefore provides a useful model to initialize the magnetic topology in upcoming MHD simulations and stellar evolution codes taking into account magneto-rotational transport processes. 
\end{abstract}

\firstsection 

\vspace*{-0.25 cm}
\section{Introduction}
The large-scale, ordered nature of magnetic fields detected at the surface of some Ap, O and B type stars and the scaling of their strengths (according to the flux conservation scenario) favour a fossil hypothesis, whose origin is not yet elucidated. To have survived since the stellar formation, a field must be stable on a dynamic (Alfv{\'e}n) timescale. It was suggested by Prendergast (1956) that a stellar magnetic field in stable axisymmetric equilibrium must contain both poloidal (meridional) and toroidal (azimuthal) components, since both are unstable on their own (Tayler 1973; Wright 1973). This was confirmed recently by numerical simulations (Braithwaite \& Spruit 2004; Braithwaite \& Nordlund 2006) showing that an arbitrary initial field evolves on an Alfv{\'e}n timescale into a stable configuration; axisymmetric mixed poloidal-toroidal fields were found. On the other hand, magnetic equilibria models displaying similar properties have been re-examined analytically by Duez \& Mathis (2010). We here address the question of their stability.

\vspace*{-0.5 cm}
\section{The model}
We deal with non force-free magnetic configurations ({\it i.e.} with a non-zero Lorentz force) in equilibrium inside a conductive fluid in absence of convection. Several reasons inclined us to focus on non force-free equilibria. First, Reisenegger (2009) reminds us that no configuration can be force-free everywhere. Although there do exist ``force-free'' configurations, they must be confined by some region or boundary layer with non-zero or singular Lorentz force. Second, non force-free equilibria have been identified in plasma physics as the result of relaxation (self-organization process involving magnetic reconnections in resistive MHD), {\it e.g.} by Montgomery \& Phillips (1988). 
Third, as shown by Duez \& Mathis (2010), this family of equilibria is a generalization of Taylor states (force-free relaxed equilibria; see Taylor 1974) in a stellar context, where the stratification of the medium plays a crucial role. The equilibrium obtained is described in detail in Duez \& Mathis (2010) as the lowest energy configuration conserving the invariants of the problem (during the relaxation phase) which are the magnetic helicity (preventing the rapid energy decay) and the mass enclosed in magnetic poloidal flux surfaces (to account for the non force-free property). 

\vspace*{-0.5 cm}
\section{Stability numerical analysis}
We use the Stagger code (Nordlund \& Galsgaard 1995). We model the star as a self-gravitating ball of ideal gas $(\gamma = 5/3)$ with radial density and pressure profiles initially obeying the polytropic relation $P \propto \rho^{1+(1/n)}$, with index n = 3, therefore stably stratified. More details on the numerical model setup can be found in Braithwaite \& Nordlund 2006. The configuration is then submitted to white perturbations ($1\%$ in density). The dynamical evolution of the mixed configuration is compared to its purely poloidal and its purely toroidal components, whose behaviour are well known to be unstable due to kink-type instabilities. The magnetic and velocity amplitudes are plotted on Fig. 1. 
\begin{figure}[ht]
\vspace*{-0.1 cm}
\begin{center}
 \includegraphics[width=0.28 \textwidth]{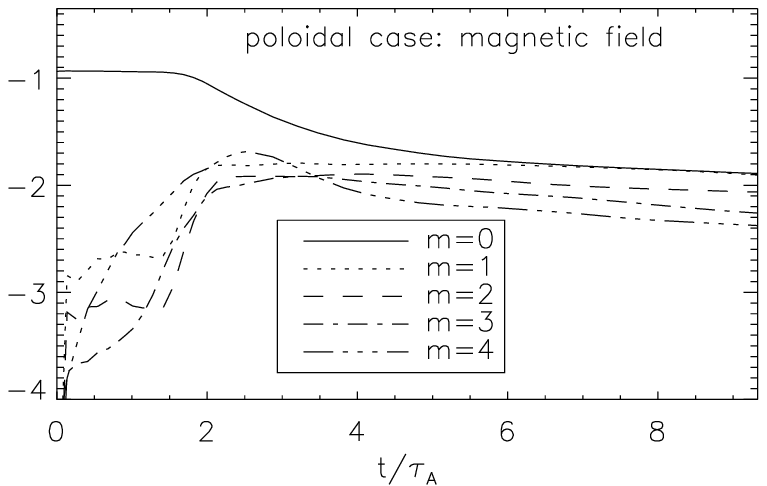}
 \includegraphics[width=0.28 \textwidth]{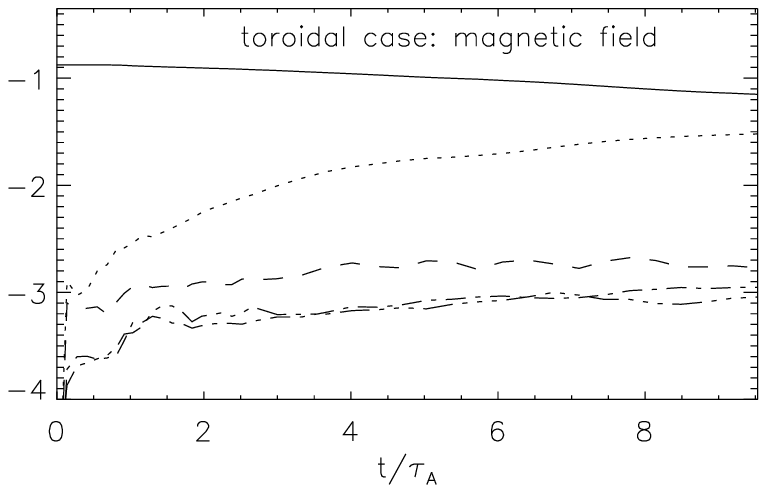}
 \includegraphics[width=0.28 \textwidth]{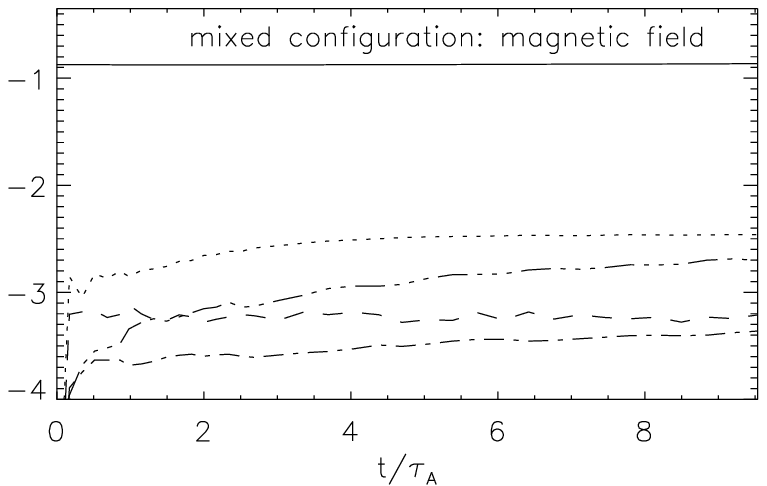}\\
 \includegraphics[width=0.28 \textwidth]{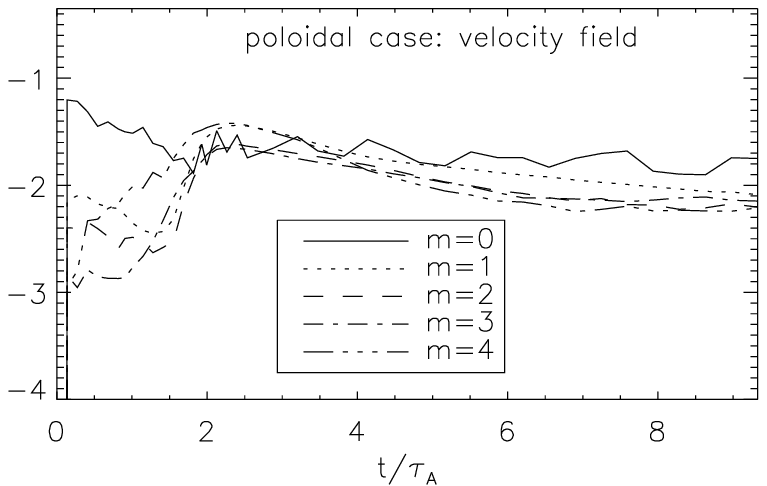}
 \includegraphics[width=0.28 \textwidth]{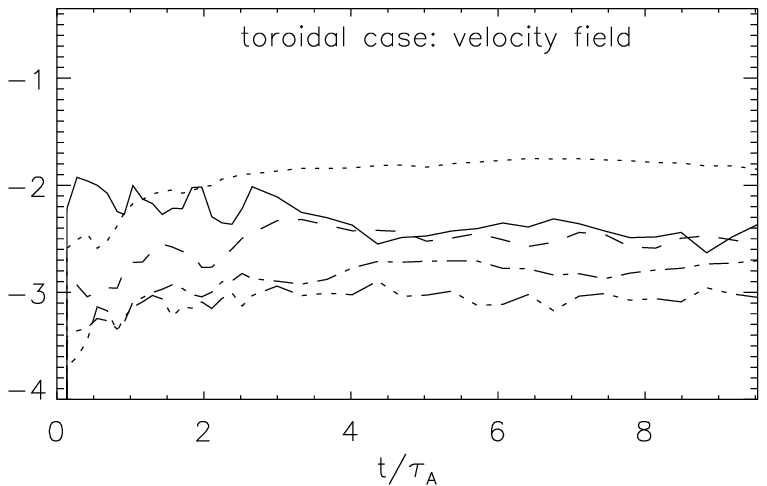}
 \includegraphics[width=0.28 \textwidth]{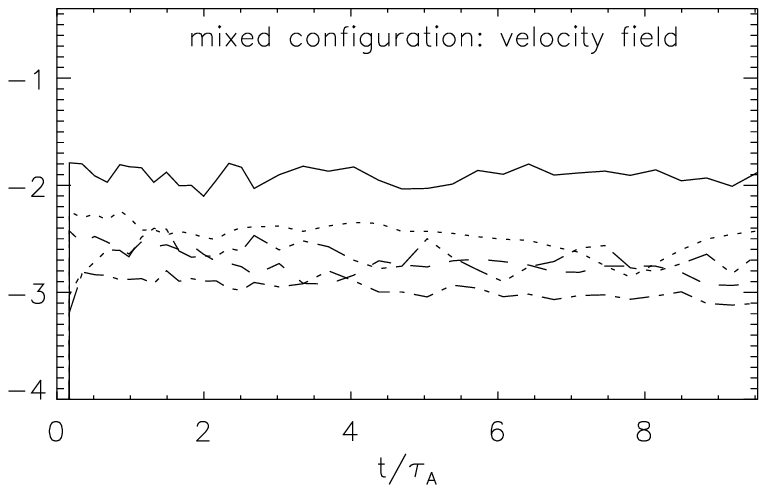}
\vspace*{-0.1 cm}
\caption{Time evolution of the (log) amplitudes in azimuthal modes $m = 0$ to $4$ averaged over the stellar volume of the magnetic field (top row) and the velocity field (bottom row) in the simulations with the purely poloidal field (left), purely toroidal field (middle) and the mixed field (right). Initially, all the magnetic energy is in the $m = 0$ mode since the initial conditions are axisymmetric.}
\label{fig1}
\end{center}
\end{figure}
As we can see, in contrast to these unstable configurations, the mixed poloidal-toroidal one does not exhibit any sign of instability, even for high azimuthal wavenumbers (up to about 40). This is the first time the stability of an analytically-derived stellar magnetic equilibrium has been confirmed numerically. This result has strong astrophysical implications: the confuiguration provides a good initial condition to magnetohydrodynamic simulations and magneto-rotational transport to be included in next generation stellar evolution codes; furthermore it could help to appreciate the internal magnetic structure of neutron stars and magnetic white dwarfs.



\vspace*{-0.5 cm}
\begin{thebibliography}{}

\bibitem[Braithwaite \& Nordlund (2006)]{Braithwaite:2006}{Braithwaite, J. \& Nordlund, \AA} 2006, \textit{A\&A}, 450,1077

\bibitem[Braithwaite \& Spruit (2004)]{Braithwaite:2004}{Braithwaite, J. \& Spruit, H.} 2004, \textit{Nature}, 431,819

\bibitem[Duez \& Mathis (2010)]{Duez:2010}{Duez, V. \& Mathis, S.} 2010, A\&A, 517, A58+

\bibitem[Montgomery (1988)]{Montgomery:1988}{Montgomery, D. \& Philips, L.} 1988, \textit{Phys. Rev. A}, 38, 2953

\bibitem[Nordlund (1995)]{Nordlund:1995}{Nordlund, \AA. \& Galsgaard, K.} 1995, 
http://www.astro.ku.dk/ aake/papers/95.ps.gz

\bibitem[Prendergast (1956)]{Prendergast:1956}{Prendergast, K. H.} 1956, \textit{ApJ}, 123, 498


\bibitem[Tayler (1973)]{Tayler:1973}{Tayler, R.J.} 1973, \textit{MNRAS}, 161, 365

\bibitem[Taylor (1974)]{Taylor:1974}{Taylor, J.B.} 1974, \textit{Phys. Rev. Lett.}, 33, 1139

\bibitem[Wright (1973)]{Wright:1973}{Wright, G. A. E.} 1973, \textit{MNRAS}, 162, 339

\end{thebibliography}
\end{document}